\documentclass[aps,prl,twocolumn,showpacs,floatfix,preprintnumbers,nofootinbib,superscriptaddress]{revtex4}

\usepackage{graphicx}
\usepackage{color}
\usepackage{natbib}
\usepackage{multirow}

\usepackage{epsfig}
\usepackage{graphicx}
\usepackage{amsfonts}
\usepackage{amssymb}
\usepackage{latexsym}
\usepackage{amsmath}
\def\ee{\end{equation}}
\def\ba{\begin{eqnarray}}
\def\ea{\end{eqnarray}}

\def\bq{\begin{quote}}
\def\eq{\end{quote}}

 at 10truept

\newcommand{\beq}{\begin{equation}}
\newcommand{\eeq}{\end{equation}}
\newcommand{\beqa}{\begin{eqnarray}}
\newcommand{\eeqa}{\end{eqnarray}}
\newcommand{\bea}{\begin{eqnarray}}
\newcommand{\eea}{\end{eqnarray}}
\newcommand{\p}{\partial}

 \newcommand{\ep}{\epsilon}

\def\lesssim{~\mbox{\raisebox{-.6ex}{$\stackrel{<}{\sim}$}}~}

\def\ltap{\ \raise.3ex\hbox{$<$\kern-.75em\lower1ex\hbox{$\sim$}}\ }
\def\gtap{\ \raise.3ex\hbox{$>$\kern-.75em\lower1ex\hbox{$\sim$}}\ }
\def\gl{\ \raise.5ex\hbox{$>$}\kern-.8em\lower.5ex\hbox{$<$}\ }
\def\roughly#1{\raise.3ex\hbox{$#1$\kern-.75em\lower1ex\hbox{$\sim$}}}

\begin{document}

\title{ Chaotic inflation with curvaton induced running} 
\author{Martin S. Sloth} 

\affiliation{CP$^3$-Origins, Center for Cosmology and Particle Physics Phenomenology \\ University of Southern Denmark, Campusvej 55, 5230 Odense M, Denmark}

\date{ }

\begin{abstract}
While dust contamination now appears as a likely explanation of the apparent tension between the recent BICEP2 data and the Planck data, we will here explore the consequences of  a large running in the spectral index as suggested by the BICEP2 collaboration as an alternative explanation of the apparent tension, but which would be in conflict with prediction of the simplest model of chaotic inflation. The large field chaotic model is sensitive to UV physics, and the nontrivial running of the spectral index suggested by the BICEP2 collaboration could therefore, if true, be telling us some additional new information about the UV completion of inflation.  However, before we would be able to draw such strong conclusions with confidence, we would first have to also carefully exclude all the alternatives.  Assuming monomial chaotic inflation is the right theory of inflation, we therefore explore the possibility that the running could be due to some other less UV sensitive degree of freedom.  As an example, we ask if it is possible that the curvature perturbation spectrum has a contribution from a curvaton, which makes up for the large running in the spectrum. We find that this effect could mask the information we can extract about the UV physics. We also study different different models, which might lead to a large negative intrinsic running of the curvaton. 
 \end{abstract}

\maketitle

\section{Introduction}

The recent data from the BICEP2 experiment has been interpreted by the BICEP2 collaboration as the discovery of primordial tensor modes, with a tensor-to-scalar ratio, $r$, measured to be $r=0.2^{+0.07}_{-0.05}$ \cite{Ade:2014xna}. While such a large value for the tensor-to-scalar ratio would fit well the predictions of chaotic inflation with a quadratic potential \cite{Linde:1983gd}, this model is in some tension with the Planck data \cite{Ade:2013zuv}. 

The source of the tension of quadratic chaotic inflation model with the Planck data is the prediction of a large value of $r\approx 0.15$ together with a small value of the running of the spectral index $\alpha$. If the assumption of a small running of the spectral index is relaxed, the value of $r$ found by BICEP2 is not tension with Planck, and assuming $r=0.2$, as found by BICEP2, the Planck data appear to imply $\alpha =-0.02$. A running of the spectral index of this size was already favored by the WMAP7+SPT data \cite{Hou:2012xq}.

While the measured value of $r$ is likely to become lower with better foreground subtraction or better understanding of other systematical issues \cite{Mortonson:2014bja,Flauger:2014qra}, it is interesting meanwhile to ask the question, what would it require from a theoretical point of view to bring BICEP2 and Planck in better agreement? Naively, the most straight forward approach would be to modify inflationary sector itself by allowing for a more nontrivial inflation potential \cite{Czerny:2014wua} or perhaps a modification of the initial state \cite{Collins:2014yua,Ashoorioon:2014nta}. If this is the correct interpretation of the data, it would be striking because in order to achieve the large tensor-to-scalar ratio in the first place, we would need a large field model of inflation, which is sensitive to UV physics. Therefore, any modification of the inflationary sector would carry information about the UV completion of inflation. However, assuming that the source of the tension between BICEP2 and Planck is not an issue of estimating systematic errors in either of experiments but actually contains physical information, one might want to exclude other possible alternatives before drawing the strong conclusion that this is new additional information about the UV completion of inflation. Therefore we will follow a simple logic. We will assume that the data is telling us that the minimal monomial chaotic inflation is the right theory of inflation, and the running is coming from some other light and less UV sensitive degree of freedom.  As a model, we will as a starting point consider the case where the curvature perturbation spectrum has a contribution from a curvaton \cite{Enqvist:2001zp,Lyth:2001nq,Moroi:2001ct}\footnote{For some earlier related work, see also \cite{Mollerach:1989hu,Linde:1996gt}.}, which makes up for the large running in the spectrum.

Models of mixed inflaton-curvaton scenarios has been consired before in the literature (see f.ex.  \cite{Langlois:2004nn,Sloth:2005yx,Kinney:2012ik,Enqvist:2013paa,Byrnes:2014xua} ). However in \cite{Kinney:2012ik} it was assumed that both the inflaton and the curvaton has a simple power law spectrum. This is a good approximation when there is no running in neither the inflaton or the curvaton spectrum, but will always lead to a positive running in the final spectrum $\alpha >0$.  Below we will therefore generalize the setup of  \cite{Kinney:2012ik} to the case where there is a large intrinsic negative running in the curvaton spectrum. 

Curvaton models with a large negative running has been discussed before in  \cite{Takahashi:2013tj,Peloso:2014oza} as a possible way of accommodating the claimed running of $\alpha = -0.024$  in the WMAP7+SPT data. But while a curvaton model in its pure form would lead to a vanishing tensor-to-scalar ratio and therefore be ruled out by BICEP2, one might speculate that it could be a component contributing to the total curvature perturbation, which however is responsible for all of the large running implied by the new data, as we will discuss below.

\section{The model}

We will assume that inflationary part is given by a model of chaotic inflation with the potential \cite{Linde:1983gd}
\beq
V(\phi) =\lambda m_{pl}^4 \left(\frac{\phi}{m_{pl}}\right)^{n}~.
\eeq
Large field models of inflation, like the model above, are protected by an approximate shift symmetry, however it is a challenge to build UV complete models that lead to this softly broken shift symmetry, since quantum gravity does not appear to respect continuous global symmetries (for a detailed review of these arguments see \cite{Kaloper:2011jz} and see \cite{Kehagias:2014wza,Hertzberg:2014aha,Lyth:2014yya} for some recent related comments). One can consider models relying on an axion where the effect of the symmetry breaking is attempted to be kept under control by the periodicity of the axion potential. However, it does not appear that a super-Planckian axion decay constant, as would be required in a large field model of inflation, is consistent with string theory \cite{Banks:2003sx}. Some model building efforts has been diverted to resolve this problem, and at present monodromy type models seems to be a promising direction \cite{Silverstein:2008sg}. It has been argued that quadratic chaotic inflation can be naturally realized from a low energy effective point of view in these type of setups \cite{Kaloper:2011jz,Kaloper:2008fb}. In addition to these ideas, it has been argued that within a pure supergravity framework, that essentially all models of chaotic inflation with $n>0$ can be constructed \cite{Kallosh:2011qk,Kallosh:2010xz,Kallosh:2010ug}, and models with $n>2$ motivated by a superconformal approach to supergravity was put forward in  \cite{Kallosh:2011qk,Kallosh:2010xz,Kallosh:2010ug}. 

The BICEP2 result for the measured value of $r$ fits with the simple quadratic model of chaotic inflation, but a large running was suggested to obtain better agreement with Planck. While it may be that the  corrections to the quadratic potential could become important, and carry nontrivial information about the UV completion, it may be important to check if other physics, separate from the UV sensitive large field inflaton potential, can affect such conclusions. 

This motivates our study of whether it is possible that the large running of the spectral index can be induced by a curvaton with sub-Planckian field values. 
%In the axiverse string landscape \cite{Arvanitaki:2009fg}, one may generically expect many axion type particles around, and it may not be completely unexpected if an axion type particle with sub-Planckian decay constant could behave as a curvaton component. 

We therefore consider the case where the scalar curvature perturbation spectrum is formed as a linear combination of two uncorrelated contributions from the the inflaton and the curvaton respectively. This can be parametrized as \cite{Kinney:2012ik}
\beq\label{P1}
P(k) = A_s \left[(1-f)\left(\frac{k}{k_*}\right)^{(n_{inf}-1)}+f\left(\frac{k}{k_*}\right)^{(n_{\sigma}(k)-1)}\right]~,
\eeq
where we have allowed for an intrinsic running in the curvaton spectrum, by making $n_{\sigma}$ scale dependent. As mentioned above, it was shown in \cite{Kinney:2012ik}, that if we do not allow for an intrinsic running of either the inflaton or the curvaton spectrum, the induced running of the total spectrum is always positive. Here will relax the assumption of ignoring the intrinsic running in the curvaton, and consider the possibility that an intrinsic running of the curvaton can induce a negative running in the final spectrum.

If the fraction of the curvaton component is small $f<<1$, the prediction of the tensor-scalar ratio from chaotic inflation remains almost intact, and would have 
\beq
r  = (1-f)16\ep
\eeq
where $\ep = n/(4 N_*)$ and $N_* \approx 55$ is e-foldings left of inflation when the observable modes exits the horizon.

The spectral index of the total curvature perturbation is given by 
\beq
\frac{d \ln P(k)}{d\ln k} = n_s -1
\eeq
and the running is defined as 
\beq
\frac{d n_s(k) }{d\ln k} = \alpha~.
\eeq
For convenience we will also define the intrinsic running of the curvaton as
\beq
\frac{d n_\sigma(k) }{d\ln k} = \alpha_\sigma~,
\eeq
while we will ignore the intrinsic running of the inflaton.

By Taylor expanding (\ref{P1}) above around some pivot scale $k_*$, we obtain
\beq
n_s-1 = n_{inf}-1 +f(n_{\sigma}-n_{inf})  ~.
\eeq
On the other hand, assuming $(n_{inf}-1)^2 \sim (n_{\sigma}-1)^2 << \alpha_{\sigma}$,  we find for the running 
\beq
\alpha \approx f \alpha_{\sigma}
\eeq
Thus with $f\approx 0.4$, then on the scale $k_*$ with $\alpha_{\sigma} \sim -0.02$, one would find $\alpha \approx -0.01$ and for chaotic inflation with the quartic potential $V(\phi) = \lambda\phi^4$ one would obtain $r\approx 0.2$. Thus these parameters naively seems to be in simultaneous agreement with both BICEP and Planck. 

The amplitude of the curvaton perturbation is 
\beq
P_{\zeta}^{\sigma} = r_{dec}^2 \frac{1}{9\pi^2}\frac{H_*^2}{\sigma_*^2}
\eeq
where $*$ denotes the time at which CMB scales left the horizon, and $r_{dec}$ is the curvaton fraction of the total energy density at time of decay. Thus in order for the curvaton perturbation to be a fraction $f$ of the total curvature perturbation as measured by Planck \cite{Ade:2013zuv}
\beq
P_{\zeta} \approx 2.2 \times10^{-9}
\eeq
with $H_* = 1.1 \times10^{14}$ GeV fixed by requiring $r=0.2$ as measured by BICEP2, we need
\beq
\sigma_* \approx  0.02 \frac{r_{dec}}{f^{1/2}} m_{pl}~.
\eeq
Thus, if we take $r_{dec} \approx 0.1$, then with  $0.1\lesssim f \leqslant1$, the curvaton field value stays two orders of magnitude below the Planck mass.

Assuming the curvaton is light, one may still worry how the curvaton gets displaced so high up in its potential initially compared with the Hubble scale. However, we note that the initial field value of the curvaton roughly corresponds to the GUT scale around $10^{16}$ GeV. Thus, it is natural to believe that a phase transition around the GUT scale might have initially displaced the curvaton in its potential. Since the observable part of inflation also happens near the GUT scale, the curvaton might have been displaced just before the last $55$ e-folds of inflation, which could explain the relatively high value for the initial curvaton field. Another possibility is that inflation is long and the dispersion of the field is given by the equilibrium value, which will imply that a typical field value is $\sigma_* \sim H^2/m$, which for $m\sim 0.01 H$ again will give the correct vacuum expectation value.

Since the curvaton component is non-Gaussian, there is also a possibility that the total curvature perturbation will have a non-Gaussian component. The non-Gaussianity can be estimated by writing
\beq
f_{NL} \simeq \frac{\left< \zeta_{\sigma}^3\right>}{\left< \zeta_{inf}^2\right>^2}  = \frac{\left< \zeta_{\sigma}^2\right>^2}{\left< \zeta_{inf}^2\right>^2} \frac{\left< \zeta_{\sigma}^3\right>}{\left< \zeta_{\sigma}^2\right>^2} 
\eeq
where it is well known that for $r_{dec} <<1$ \cite{Lyth:2002my}
\beq
\frac{\left< \zeta_{\sigma}^3\right>}{\left< \zeta_{\sigma}^2\right>^2} =\frac{5}{4r_{dec}}~,
\eeq
and one will obtain
\beq\label{fNL1}
f_{NL} = \frac{5}{4}\frac{f^2}{r_{dec}}~.
\eeq
Thus, with the numerical examples $0.1<f <0.5$ and $r_{dec}\approx 0.1$, we obtain $0.1< f_{NL} < 2$ consistent with current upper bounds.

As a consequence of the above observations, it is interesting to consider if there are curvaton models, which self-consistently can lead to a large intrinsic negative running. In the light of the large negative running  of the spectral index previously indicated by the SPT data \cite{Hou:2012xq}, models of this type has already been explored to some extend in the literature \cite{Takahashi:2013tj,Peloso:2014oza}. Below we will explore some new aspects of how one can achieve a large intrinsic negative running.

\section{Curvaton with large negative intrinsic running }

In order to illustrate in a simplified manner how one can have a large negative intrinsic running of the curvaton,  let us consider for a moment the limit where we can treat the Hubble rate as a constant and take $\epsilon<< V''/H^2$,  where $V'' =\p^2 V/\p\sigma^2$. If we have the situation where the field, $\sigma$, is only gravitationally coupled to other fields, then in this approximation we have 
\bea
n_\sigma-1 &\approx&  \frac{2}{3}\frac{V''}{H^2}\\
\alpha_\sigma &\approx& -\frac{2}{9}\frac{V'V'''}{H^4}
\eea
where for a simple polynomial potential we have $V' V''' \propto V''^2$, and thus the running will be $\alpha_\sigma \propto (n_\sigma-1)^2$, and therefore it is hard to achieve a large running in this simple case. Relaxing the assumption $\epsilon<< V''/H^2$ complicates the argument slightly, but does not change this conclusion. Below we will therefore consider two different ways to circumvent the assumptions that led to this pessimistic conclusion. First we will consider the case where the potential has an explicit dependence on a second fast-rolling scalar field, and by proxy this induce a large time dependence in the mass of $\sigma$, which leads to a large running of its spectral index. Second we consider the case where the potential is not on a polynomial form, but periodic instead.

\section{Example 1: Running by proxy }

Here we will explore the idea that a large intrinsic running of the curvaton could be obtained by making the curvaton mass depend on a fast-rolling proxy field. Let us assume that there is an additional test scalar field, $\chi$, present during inflation. We will then assume that the potential, $V(\sigma)$, is having a logarithmic correction from the additional scalar given by
 \beq
 V(\sigma) = \frac{1}{2}m_\sigma^2\sigma^2 - \frac{1}{2}\Delta m_\sigma^2 \log(\chi/\chi_0) \sigma^2~,
 \eeq
and the potential for $\chi$ is 
\beq
V(\chi) = V_0-\frac{1}{2}m_\chi^2 \chi^2~.
\eeq
Assuming that $\Delta m_\sigma^2 \log(\chi/\chi_0) \sigma^2<< m_\chi^2 \chi^2$, the equation of motion for the additional scalar is 
\beq
\ddot \chi +3H\dot\chi = m_\chi^2 \chi~.
\eeq
Following \cite{Linde:2001ae}, one can assume $\chi=\chi_0 e^{i\omega t}$, which implies $\omega = i(3H/2\pm \sqrt{9H^2/4 +m_\chi^2})$, which gives the fast-roll solution
\beq
\chi = \chi_0 e^{F(m_\chi/H)Ht }~,
\eeq
where $F(m^2_\chi/H^2) = \sqrt{9/4+m_\chi^2/H^2}-3/2$ and in the limit $m_\chi >> H$, $F(m_\chi^2/H^2)\approx m_\chi/H$.
Assuming also for simplicity also $\Delta m_\sigma^2 \log(\chi/\chi_0)  << m_\sigma^2$, we have that the spectral index for the $\sigma$ field is
\bea
& &n_\sigma-1= -2\ep + \frac{2}{3}\frac{V''}{H^2}\\
& &=-2\ep +\frac{2}{3}\frac{m_\sigma^2}{H^2}\left(1-\frac{\Delta m_\sigma^2}{m^2_\sigma}\log(\chi/\chi_*)\right) \approx -2\ep +\frac{2}{3}\frac{m_\sigma^2}{H^2}~.\nonumber
\eea
Now using that $d\log k = H dt$ and time derivatives of slow-roll parameters are order slow-roll parameters squared, we have 
\beq
\alpha_\sigma = \frac{d n_\sigma}{d\log k} \approx -F \frac{\Delta m_\sigma^2}{H^2}~.
\eeq

Since we assumed that $\chi$ is in the fast-roll regime $m_\chi >> H$ we can easily have $\alpha_\sigma \approx -0.02$ even when $\Delta m_\sigma^2  << m_\sigma^2<<H^2$. Since both the curvaton, $\sigma$, and the proxy field, $\chi$, gives a subdominant contribution to the total energy density, we can for simplicity assume $\rho_\sigma\approx \rho_\chi << \rho_\phi$. This implies that $m_\chi^2\chi^2 \approx m_\sigma^2\sigma^2$, so the constraint $\Delta m_\sigma^2 \log(\chi/\chi_0) \sigma^2<< m_\chi^2 \chi^2$ will also automatically be satisfied if $\Delta m_\sigma^2 \log(\chi/\chi_0)  << m_\sigma^2$ is satisfied.

However, the challenge is to keep the proxy field in fast-roll for sufficiently many e-folds. In order to have a constant running during the approximate $8$ e-folds of inflation when the CMB modes exit the horizon, we need the proxy field to keep fast rolling during $8$ e-folds. The number of e-folds during which the proxy field is fast-rolling is given by
\beq
N  = \frac{1}{F} \log\left(\frac{\chi_*}{\chi_0}\right)
\eeq 
where $\chi_0$ is the initial value of $\chi$, and $\chi_*$ is the final value of $\chi$ at the end of the fast-roll regime. Since the fluctuations of the proxy field is of order $\delta\chi \simeq m_\chi/2\pi$, it was estimated \cite{Linde:2001ae} it was estimated that the smallest possible value if $\chi_0 \simeq m_\chi/10$. Similarly, if we assume that the potential energy of $\chi$ is positive definite, then the fast-roll approximation will have to breakdown when $(1/2)m_\chi^2\chi^2 \sim V_0$, which implies $\chi_* \sim \sqrt{V_0}/m_\chi$. Using $V_0 << H^2$, we then have 
\beq
N\lesssim \frac{1}{F}\log\left(\frac{1}{H}\frac{H^2}{m_\chi^2}\right)~.
\eeq
Now it is clear that in the limit $m_\chi>>H$ where we have $F\approx m_\chi/H$,  we have that $H$ needs to be exponentially small for $N >1$ to be allowed. So while this limit could work for generating a large negative running in low scale inflation, it will not work if the BICEP2 result is correct and $H\approx 10^{-4}$. In this case we need to consider instead more moderate situations with $m_\chi \sim H$. We will for illustration consider two numerical examples. We will consider the cases where $m_\chi/H =1$ and $m_\chi^2/H^2 =10$. With $F^{-1}(1)=3.3$ we have
\beq
N\lesssim 3.3 \log\left(\frac{1}{H}\right) \approx 30~, 
\eeq
and with  $F^{-1}(10) =0.5$ we have
\beq
N\lesssim 0.5 \log\left(\frac{0.1}{H}\right) \approx 3.5~, 
\eeq
where we used $H\approx 10^{-4}$. Thus with $m_\chi \approx H$, we can sustain the running of a sufficient amount of e-folds, even with $H\sim 10^{-4}$, but if the BICEP2 result turns out to be incorrect and $H$ is very low, then we can sustain a large running even with $m_\chi>> H$.

Assuming that $F\approx 1$, then $\alpha =-0.02$ will require $\Delta m_\sigma^2 / H^2 \approx 10^{-2}$. At the end of the fast-roll regime $\log(\chi_*/\chi_0) \lesssim \log(H/m_\chi^2)$, so the requirement $m_\sigma^2 \gtrsim \delta m_\sigma^2 \log(\chi_*/\chi_0)$ implies for  $1<m_\chi^2/H^2 < 10$ and $H=10^{-4}$ that $0.1\gtrsim m_\sigma^2/H^2 \gtrsim 0.07$ which appears to be on the borderline of the acceptable in order not to destroy the flatness of the potential.

\section{Example 2: curvaton as an axion}

In \cite{Takahashi:2013tj} a modulation of the curvaton potential on the following form was considered
\beq
V(\sigma) = V_0(\sigma) +\delta V(\sigma)
\eeq
where the modulation of the curvaton potential is given by
\beq
\delta V(\sigma) = \Lambda(1-\cos(\sigma/f_\sigma))~.
\eeq
Since the field value of the curvaton, $\sigma_*$, is at least  two orders of magnitude below the Planck scale, even $V_0(\sigma)$ could be taken to have the periodic form $V_0(\sigma) = \Lambda_0(1-\cos(\sigma/\tilde f_\sigma))$ with a sub-Planckian decay constant $f_{\sigma}<<\tilde f_{\sigma}<< m_{pl}$. As discussed in \cite{Takahashi:2013tj}, such a potential could be generated by an axion\footnote{Some earlier examples where the string axion was considered as a possible curvaton candidate are \cite{Enqvist:2001zp,Dimopoulos:2003az}.} coupling to two different gauge fields, or by a complex scalar field with an approximate global $U(1)$ symmetry that gets spontaneously broken in the context of SUSY with discrete $R$ symmetry.   For generality we will here keep the potential $V_0$ unspecified, and using the equation of motion for the curvaton 
\beq
3 H \dot\sigma \simeq -V_0' (\sigma)
\eeq
it was shown that one can express the axion decay constant, $f_{\sigma}$, in terms of number of e-foldings, $\Delta N$, giving one period of the modulation induced by $\delta V(\sigma)$, and $V_0'(\sigma_*)$,
\beq\label{fsigma}
f_{\sigma} = \frac{1}{2\pi} \frac{V_0'(\sigma_*)}{3 H_*^2} \Delta N ~.
\eeq
We note that for a small curvaton mass $m_{\sigma}^2\equiv V_0''  << H_*^2$ and using  $V_0'(\sigma_*)  \sim V_0'' (\sigma_*)\sigma_*=m_{\sigma}^2\sigma_* $, the axion decay constant is indeed small $f_\sigma << m_{pl}$.

Using from\footnote{In ref. \cite{Takahashi:2013tj} the approximation of a constant Hubble rate, $H$, during inflation was used, and it was pointed out that for large field models of inflation the change in $H$ during inflation could change these estimates. However, since the change in $H$ is still small over the few e-foldings when the observable modes exits the horizon, we will assume that this effect will only be a subleasing correction.}  \cite{Takahashi:2013tj}
\beq\label{alpha}
\alpha_{\sigma}\sim \frac{2\pi}{\Delta N}(n_{\sigma}-1)
\eeq
where $\Delta N/2 > 8$ is needed to a maintain a relatively constant running over the CMB scales, one obtains by taking $n_{\sigma}-1 \simeq 0.05$ and $\Delta N \approx 16$
\beq
\alpha_\sigma \approx -0.02~.
\eeq
Since we have $\alpha = f \alpha_{\sigma}$, then in order to have a large $\alpha$, we do not want to choose too small a curvaton fraction. A large curvaton fraction however means decreased value of $r$, so to keep $r$ large enough to match BICEP, we might want to take as an example a quartic potential, $n=4$, with a $50$\% curvaton mix, $f=0.5$, which gives
\beq
\alpha \approx -0.01~, \qquad r = 0.15  
\eeq
while a slightly smaller curvaton mix would give slightly larger $r$, but also slightly less running. For instance a $40$\% curvaton mix, $f=0.4$ would give $\alpha\approx -0.008$ but $r=0.2$. Likewise running can be increased by lowering $r$ further from $r=0.15$. 

\subsection{Constraints from gauge field production}

While we have shown that the non-Gaussianity computed in (\ref{fNL1}) can be assumed to be small with a suitable choice of parameters, there is an additional new potential source of non-Gaussianity in the axion curvaton model. The axion will typically have a $CP$ violating coupling to gauge fields of the form
\beq \label{Lint}
\mathcal{L}_{int} = \lambda \frac{\sigma}{f_{\sigma}}F\tilde F~, \qquad \mathcal{L}_{int} = \lambda \frac{\sigma}{\tilde f_{\sigma}}G\tilde G
\eeq
where $F$ and $G$ are the field strengths of two different gauge groups. In the case where inflaton itself is assumed to be an axion with similar couplings to gauge fields, it has been studied in a series of papers how a resonant production of gauge fields can lead to large non-Gaussian features inconsistent with observations. Here will discuss the similar constraints in the curvaton case, and show that a similar effect can possibly lead to a new source of non-Guassianity in the axionic curvaton case.

In order to have a resonant production of gauge fields, the curvaton must be relative strongly coupled to the gauge fields. Let us therefore ignore the weaker coupling given by $\tilde f_\sigma$ and consider for a moment the effective Lagrangian
\beq
\mathcal{L}_{eff} = -\frac{1}{2}(\p\sigma)^2-V(\sigma)-\frac{1}{4}F_{\mu\nu}F^{\mu\nu}-  \lambda \frac{\sigma}{f_{\sigma}}F_{\mu\nu}\tilde F_{\mu\nu}~.
\eeq
Introducing the vector potential $A_{\mu}$ and choosing the Coulomb gauge with $A_0$, the equation of motion for the spatial part of the vector potential is
\beq
\left(\frac{\p^2}{\p\tau^2} -\nabla^2-\lambda\frac{\sigma'}{f_{\sigma}}\vec{\nabla}\times\right)\vec{A} =0~,\qquad \vec{\nabla}\cdot \vec{A}=0
\eeq 
where $\tau$ is conformal time and prime denotes derivatives with respect to conformal time. Now expanding the vector field into annihilation and creation operators
\beq
\hat{\vec{A}} =\sum_{\rho=\pm} \int \frac{d^3 k}{(2\pi)^{3/2}}\left[\vec\ep_\lambda(\vec{k})A_\rho(\tau,\vec{k})a^\rho_{\vec k} e^{i \vec{k}\cdot \vec x}+\textrm{h.c.}\right]
\eeq
and defining the polarization vectors such that $\vec k \cdot \vec \ep_\pm =0$, $\vec k\times  \vec \ep_\pm =\mp i|\vec k | \vec \ep_\pm $, the equation of motion for the two polarizations of the vector field takes the form
\beq
\frac{d^2}{d\tau^2} A_\pm(\tau,k) +\left[k^2\pm 2k\frac{\xi}{\tau}\right]A_\pm(\tau,k)=0~,
\eeq
where one conventionally defines
\beq
\xi\equiv \frac{\lambda\dot\sigma}{2 f_\sigma H}
\eeq
in the approximation where $\dot\sigma\approx \textrm{constant}$.

In the case where $\xi \gtrsim 1$ it has previously been demonstrated that the $A_+$ mode will be resonantly amplified when $1/(8\xi)\lesssim -k\tau \lesssim 2\xi$, and one finds in this regime \cite{Anber:2009ua}
\beq\label{A+}
A_+ (\tau,k) \approx \frac{1}{\sqrt{2k}}\left(\frac{k}{2\xi a H}\right)^{1/4} e^{\pi \xi -2\sqrt{2\xi k/aH}}
\eeq
which shows an exponential enhancement of $A_+$ by the factor $e^{\pi\xi}$, while $A_-$ is not amplified and can be ignored. 

Since the axion couples directly to the gauge field through the coupling (\ref{Lint}), the resonant amplification of the gauge field can affect the correlation functions of the axion at the loop level and source a significant non-Gaussian component, as we will discuss below.

Let us consider the perturbation of the axion around its homogenous background, and write $\sigma(\tau,\vec x) \equiv \sigma(\tau) + \delta\sigma(\tau,\vec x)$. Then the equations of motion for the background become
\bea
 &&\ddot\sigma +3 H\dot\sigma +V'(\sigma) = \frac{\lambda}{f} \left< \vec{E}\cdot \vec{B}\right>\nonumber\\
&&3 H^2 =\frac{1}{2}\dot\phi^2 +V(\phi)+\frac{1}{2}\left<\vec{E}^2+\vec{B}^2\right>
\eea
where $\vec{B}=a^{-2}\vec\nabla\times\vec A$ and $E=-a^{-2}\vec A'$. Using the solution (\ref{A+}) one can show that $ \left< \vec{E}\cdot \vec{B}\right> \simeq -2.4 \cdot 10^{-4}(H/\xi)^4\exp(2\pi\xi)$ and $\frac{1}{2}\left<\vec{E}^2+\vec{B}^2\right> \simeq 1.4\cdot 10^{-4}(H^4/\xi^3)\exp(2\pi\xi)$. This implies that we expect the effect of the gauge fields on the background evolution to be negligible only when \cite{Anber:2009ua}
\bea
&&\frac{H^2}{2\pi |\dot\sigma |}<< 13 \xi^{3/2}e^{-\pi\xi}\nonumber\\
&&H<< 146\xi^{3/2}e^{-\pi\xi}
\eea
where for $H\simeq 10^{-4}$ the second constraint implies $\xi < 5.3$. Now let us consider the first constraint above within the same parameter region. Assuming backreaction is small then we have for $m_\sigma<<H$ that $3H\dot\sigma = V'(\sigma) \approx \Lambda_0/\tilde f_\sigma \sin(\sigma/\tilde f_\sigma) <  \Lambda_0/\tilde f_\sigma$ it follows that for the choice  $m_\sigma/H \lesssim10^{-2}$ and $H\sim10^{-4}$, then $H^2/(2\pi |\dot\sigma |) \gtrsim 1$, and the first constraint is never satisfied for any $\xi\gtrsim 1$. Thus, in order to avoid backreaction issues we will need to take 
\beq \label{bc}
\xi <1
\eeq
such that the gauge fields are not enhanced.

The backraction constraint (\ref{bc}) is equivalent to 
\beq
f_\sigma\gtrsim \frac{\lambda}{2}\frac{\dot\sigma}{H} 
\eeq
which when combined with (\ref{fsigma}), and using again $3H\dot\sigma = V'(\sigma)$, gives
\beq
\lambda\lesssim \frac{\Delta N}{\pi}~.
\eeq
Thus in models where $\lambda$ is large (for such models see appendix B of  \cite{Anber:2009ua}), one will generically have backreaction issues which would invalidate the analysis of \cite{Takahashi:2013tj} leading to (\ref{alpha}). On the other hand it is not a problem from a fundamental point of view to have $\lambda <1$ (which may in fact be theoretically preferred). Finally one should also mention the possibility that the background dynamics of the curvaton is determined by the backreaction effect of the gauge field itself, which in the case of the inflaton is known to able to provide a slow-roll solution. We expect the same would in principle be the case here, although it is beyond the scope of the present paper to analyze this situation in details.

Let us finally consider the generalization of the backreaction constraint on the axion as a curvaton to the case where parts or all of the BICEP2 signal can be explained by dust as indicated by \cite{Mortonson:2014bja,Flauger:2014qra}. In this case we obtain instead the more general lower bound  $H^2/(2\pi |\dot\sigma |) \gtrsim 10^{4}H $, while $13 \xi^{3/2}e^{-\pi\xi}> 10^{-2}$ for moderate values  $1<\xi<3$. Thus only a slight decrease of the Hubble rate during inflation will allow for interesting values of $1<\xi<3$. For these values of $\xi$ the effect of the gauge field production on the perturbations can be relevant. To see this, we note that the equation for the axion perturbation takes the form
\bea
&&\left(\frac{\p^2}{\p\tau^2}+2\frac{a'}{a}\frac{\p}{\p\tau}-\nabla^2+a^2V''\right)\delta\sigma=\nonumber\\
&&\qquad \qquad a^2\frac{\lambda}{f_\sigma}\left( \vec{E}\cdot \vec{B}-\left< \vec{E}\cdot \vec{B}\right>\right)
\eea
which was solved in \cite{Barnaby:2010vf,Barnaby:2011vw}  by writing the solution as a sum of the approximately Gaussian homogenous solution, and a non-Gaussian particular solution quadratic in the Gaussian gauge field fluctuations
\beq
\delta\sigma(\tau,\vec{x}) = \delta\sigma^{\textrm{hom.}}(\tau,\vec{x}) +\delta\sigma^{\textrm{part.}}(\tau,\vec{x})  ~.
\eeq
The calculation of $\delta\sigma_{\textrm{part.}}(\tau,\vec{x})$ follows identically to the inflate case, and we therefore omit the details of the calculation, but just give the result obtained in \cite{Barnaby:2010vf,Barnaby:2011vw} 
\bea
\left< \delta\sigma_k\delta\sigma_{k'}\right>& =&  \left< \delta\sigma^{\textrm{hom.}}_k\delta\sigma^{\textrm{hom.}}_{k'}\right>+  \left< \delta\sigma^{\textrm{part.}}_k\delta\sigma^{\textrm{part.}}_{k'}\right>\nonumber\\
&=& P_{\delta\sigma}(k)\left(1+ \frac{H^2}{2\pi|\dot\sigma|}f_2(\xi)e^{4\pi\xi}\right)
\eea
where to good approximation $f_2(\xi) = 7.5\cdot 10^{-5}/\xi^6$. 

From the lower bound  $H^2/2\pi|\dot\sigma| > (3/2\pi)H(H/m_\sigma)^2 $, we can obtain lower bound on the non-Gaussian contribution to the curvaton perturbation form the gauge field production as a function of $H$, $m_\sigma$ and $\xi$. For illustration, let us consider an example with $H=10^{-8}$, such that the backreaction constraint can easily be satisfied for moderate values of $\xi>1$, and let us also assume like before $m_\sigma \approx 0.01 H$. Then we have $H^2/2\pi|\dot\sigma| \gtrsim 10^{-4}$ and the non-Gaussian correction becomes important already for $\xi \approx 1.5$. Using now that 
\beq
\xi \simeq  \frac{\lambda}{6}\frac{m_\sigma^2}{H^2} \frac{\tilde f_\sigma}{f_\sigma }
\eeq
where we assumed like in \cite{Takahashi:2013tj} that $\sigma \approx \tilde f_\sigma$. Using also from \cite{Takahashi:2013tj}  that  $\tilde f_\sigma/f_\sigma = 10^4$, we see that  for this choice of parameters, only for $\lambda > 9$ there is significant non-Gaussian contribution, even if the scale of inflation is very low. It was shown in \cite{Barnaby:2011vw} that this contribution will have nearly equilateral shape, which makes it an interesting example of a curvaton model with large non-Gaussianity in a different shape than the squeezed limit, which is the expected shape from the usual contribution from (\ref{fNL1}).

\section{Conclusion}

Since large field models are sensitive to UV physics, the nontrivial running of the spectral index suggested by the BICEP2 collaboration could be telling us something new about the UV completion of inflation, if this interpretation of the data should turn out to hold up in the future.  However, before drawing such strong conclusions, we explored the possibility that the curvature perturbation spectrum has a contribution from a curvaton, which makes up for the large running in the spectrum. This could relax the tension of chaotic inflation with the Planck data while still be in accordance with the BICEP2 result, and it could be accommodated by a new degree of freedom, which is less sensitive to the UV physics compared to accommodating the running by a direct modification of the inflaton potential. It turns out that the possibility of a curvaton component might actually mask the information we can extract about UV physics by allowing $\lambda\phi^4$ in addition to $m^2\phi^2$ as a monomial chaotic inflation model consistent with the data. On the other hand a non-Gaussian component consistent with observations, but with $f_{NL}$ of order one, would be a feature of such models and enable us to discriminate between single field model with negative running and the model presented here, where negative running comes from the intrinsic running of the curvaton component. We further explored in more details two example mechanisms of how a large negative intrinsic running of the curvaton might arise. In the first model the mass of the curvaton varies in time due to a third fast rolling proxy field, which leads to a large running of the spectral index of the curvaton through its $\eta$ parameter. The second model we studied, is a model known already in the literature. In this model the curvaton is an axion and the periodic nature of the potential allows for the large negative intrinsic running \cite{Takahashi:2013tj}. We checked if backreaction from gauge field production could be a problem in this model, but as long as the dimensionless parameter $\lambda$ is less than one (as expected in the most simplest cases) or if the scale of inflation is lowered, there is no problem with backreaction. For completeness we then discussed the possibility of having a new signature of non-Gaussinity with equilateral shape in the curvaton model when the scale of inflation is lowered, and the BICEP2 result is not explained by primordial tensor modes.

\section*{Acknowledgements}

I would like to thank N. Kaloper, S. Nurmi and A. Riotto for discussions. The author is supported by a Jr. Group Leader Fellowship from the Lundbeck Foundation. The CP$^3$-Origins Center is partially funded by the Danish National Research Foundation, Grant No. DNRF90.

\end{document}